\documentclass[article,12pt]{article}
\usepackage[english]{babel}
\usepackage{epsfig}
\begin{document}
\large

\textbf{Relation Between the Thickness of Stellar Disks and the
Relative Mass of Dark Halo in Galaxies} \normalsize

$$
$$

A.~V. Zasov (zasov@sai.msu.ru,Sternberg Astronomical Institute,
Moscow, Russia)

D.~V.~Bizyaev (dmbiz@sai.msu.ru, Sternberg Astronomical Institute)

D.~I.~Makarov (dim@sao.ru, Special Astrophysical Observatory,
Russian Academy of Sciences, Nizhnii Arkhyz, Karachai-Cherkessia,
Russia)

N.~V.Tyurina (tiurina@sai.msu.ru, Sternberg Astronomical
Institute)

\begin{abstract}

We consider a thickness of stellar disks of late-type galaxies by
analyzing the $R­$ and $K_s$ band photometric profiles for two
independent samples of edge-on galaxies. The main goal is to
verify a hypotesis that a thickness of old stellar disks is
related to the relative masses of the spherical and disk
components of galaxies. We confirm that the radial-to-vertical
scale length ratio for galactic disks increases (the disks become
thinner) with the increasing of total mass-to-light ratio of the
galaxies, which characterize the contribution of dark halo to the
total mass, and with the decreasing of central deprojected disk
brightness (surface density). Our results are in good agreement
with numerical models of collisionless disks evolved from
subcritical velocity dispersion state to a marginally stable
equilibrium state. This suggests that in most galaxies the
vertical stellar velocity dispersion, which determine the
equilibrium disk thickness, is close to the minimum value, that
ensures disk stability. The thinnest edge-on disks appear to be
low brightness galaxies (after deprojection) in which a dark halo
mass far exceeds a mass of the stellar disk.

\end{abstract}
$$
$$

\textbf{INTRODUCTION}

Galaxy disks are complex structural components that include the
bulk of the stellar mass in most of the spiral galaxies. Their
masses and internal structure are crucial factors that determine
all large-scale active processes in galaxies, such as propagation
of density waves, star formation, and the associated phenomena.

As a rule, the brightness (and, consequently, the surface
density) of disks at large galactocentric distances $R$ decreases
with increasing $R$ following the exponential law with a radial
scale length $h$ of the order of several kpc. Another geometrical
parameter of a stellar disk -- its thickness -- can be
characterized by vertical scale length $z_0$. In an isothermal
disk the decrease of density with the distance from the galactic
plane can be described by the simple law:

\begin{equation}
\rho (z) = \rho_0 \mathrm{sech}^2({z}/{z_0}),
\end{equation}

although some other alternative approximations are possible such
as exponential or $\mathrm{sech}(z)$-model of brightness decrease
(de Grijs and van der Kruit 1996) .

The thickness $z_0$, or the vertical disk scale height, is
primarily determined by the local disk density and stellar
velocity dispersion. However, as we can see in our own Galaxy,
young and old stars have different velocity dispersions resulting
in a rather complex vertical disk structure. Actually, since the
bulk of the disk mass in spiral galaxies consists of stars that
are several billion years old, hereafter we assign the
photometrically determined thickness to the old stellar disk.
Note that the insignificant color gradients in edge-on galaxies in
the direction perpendicular to the disk plane beyond the narrow
dust lane along the major axis (see  de Grijs 1998 and references
therein) are indicative of a rather homogeneous stellar content
of old disks.

In contrast to radial scale length $h$, the disk thickness can be
measured directly only in galaxies where disks are seen edge-on.
The relative disk thickness can be characterized, to a first
approximation, by the outer isophotal axial ratios $b/a$ of these
galaxies, although inferring the  vertical-to-horizontal scale
length ratio $z_0/h$ from photometric data requires modeling the
3D luminosity distribution of the disk (to make corrections for
projection effect).

The observed brightness distribution $\mu(r,z)$ (determined
neglecting absorption) in a finite-thickness isothermal disk seen
edge-on is related to the parameters $h$ and $z_0$ via modified
Bessel's functions of the first kind, $K_1(r/h)$ (van der Kruit
and Searle 1981a):

\begin{equation}
\mu(r,z) =
\mu(0,0)\mathrm{sech}^2\left(\frac{z}{z_0}\right)\frac{r}{h}K_1\left(\frac{r}{h}\right),
\label{e:main}
\end{equation}

where $r$ and $z$ are the sky-plane coordinates. Given relation
(2),  $z_0$ and $h$ can be determined from vertical and
major-axis photometric cross sections, respectively. At the
peripheral regions of galaxies of great importance may be such
parameter  as the so-called disk break or cutoff radius, $R_c$,
beyond which the decrease of the disk brightness is described by
a shorter exponential scale length than at smaller galactocentric
distances. According to different estimates, $R_c$ is typically
equal to 3--5 $h$ (see de Grijs et al. 2001; de Grijs and van der
Kruit 1996 and references therein).

Indirect estimates of $z_0$ could  be obtained if the stellar
velocity dispersion of the  old disk population were known.
However, such estimates require certain assumptions about the
surface density or integrated mass of the exponential disk
(Bottema 1993). The reverse is also true: given the disk
thickness, velocity dispersion measurements make it possible to
estimate the local surface brightness of the disk and,
consequently, its total mass.

The observations of edge-on galaxies showed that the disk
thickness varies over a wide range from one galaxy to another,
and the apparent axial ratio can be as high as 10--20 for the
thinnest disks (Kudrya et al. 1994; Karachentsev et al. 1997)
What determines the relative disk thickness remains an open
question. It appears to correlate with morphological type,
although the latter is determined rather uncertainly for edge-on
galaxies  -- it is inferred not from the shape of spirals but
only from the relative size and luminosity of the bulge. The
disks of late-type galaxies  (Sc--Sd) are, on the average,
"thinner" than those of early-type objects (Karachentsev et al.
1997; de Grijs 1998; Ma et al. 1997, 1999). According to de Grijs
(1998), the $h/z_0$ ratio in his sample of edge-on galaxies
varies from 1.5--2 for early-type spirals to 3--8 for Sc--Sd
galaxies. However, the relative disk thickness does not show any
direct correlation with the rotation velocity or luminosity. To
illustrate these conclusions, in Fig.~1 we compare the ($B$-band)
$a/b$ ratio according to the Flat Galaxy Catalog (RFGC)
(Karachentsev et al. 1999) with the known HI line width
($W_{50}$), which is approximately equal to twice the maximum
velocity $V$ of disk rotation, and with absolute magnitude, $M_B$
(both parameters adopted from LEDA catalog).

\begin{figure}
\epsfxsize=9cm {\epsfbox{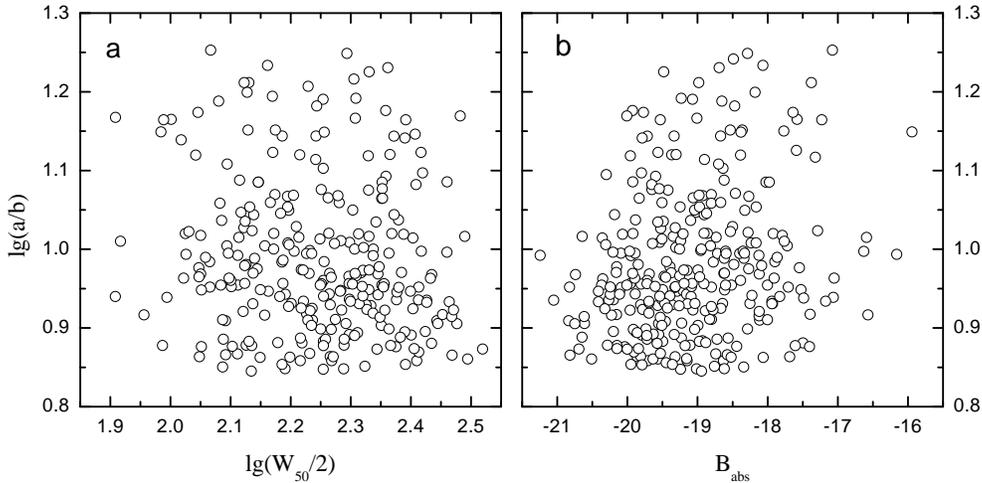}} \caption{Diagrams illustrating
the absence of correlation between the observed axial ratio $a/b$
and the HI line halfwidth, $W_{50}$/2 (a) or absolute magnitude
$B_{\mathrm{abs}}$ (b) for 340 galaxies from RFGC catalog
(Karachentsev et al. 1999). \hfill}
\end{figure}


It can be expected, however, from the most general considerations
that the relative thickness of the equilibrium disk  (at least
its minimum possible value) must reflect its kinematic
characteristics. The disk thickness at a given $R$ is indeed
determined by its local surface density and local dispersion
$C_z$ of stellar velocities in the direction perpendicular to the
disk plane. On the other hand, $C_z$ and radial dispersion $C_r$
are interrelated quantities \footnote{According to Gerssen et al.
(2000), direct estimates obtained for several galaxies yield
$C_z/C_r \approx0.5-0.8$; within approximately the same interval
(0.35--0.8) fall the ratios obtained by numerical simulation of
the dynamical evolution of initially "cold" collisionless disks
(Mikhailova et al. 2001). The condition of stability against
bending perturbations for collisionless disk yields $C_z/C_r
\approx0.37$(Polyachenko and Shukhman 1977)} with the minimum
$C_r$ determined by the condition of local gravitational
stability of the disk. Zasov et al. (1991) argued that if the
radial dispersion $C_r$ of stellar velocities in an old stellar
disk is close or proportional to the critical threshold for
gravitational (Jeans) instability of the rotating disk, and
velocity dispersion $C_z$ along the $z$-coordinate is
proportional to $C_r$, then the relative disk thickness should
increase with decreasing relative mass of the galactic halo.

Indeed, to a first approximation (neglecting the $z$-component of
the acceleration due to the spherical component of the galaxy),
$z_0\approx C_z^2 / \pi G\sigma$   (here $\sigma$ is the disk
surface density). Let radial velocity dispersion be equal to $C_r
= Q\times3.36G\sigma / \kappa$, where $\kappa \sim V/R$ is the
epicyclic frequency and the Toomre parameter $Q = 1$ corresponds
to a thin uniform disk that is marginally stable (in Toomre's
sense)
 to radial perturbations. In general case, $Q$ is a
function of radial distance $R$. Beyond the central
bulge-dominated region it varies slowly with $R$ gradually
increasing toward the periphery (Bottema 1993). However numerical
models of marginally stable disks show that parameter $Q$ remains
almost constant over a wide $R$ interval beyond the central
region and its value ($Q \approx 1.2-1.5$ between 1 and 2 radial
scalelengths $h$) depends only slightly on the mass of the
spherical and disk components of a galaxy or the shape of its
rotation curve (Khoperskov et al. 2002). Using simplified
relationships for $z_0$ and $C_r$ given above, and taking $C_z /
C_r $ and $Q(R) \approx \mathrm{const}$, one may obtain  that the
vertical-to-radial disk scale length ratio can be easily
expressed in terms of other parameters ratios:

\begin{equation}
\frac{z_0}{h} \sim \frac{C_z^2}{\sigma h} \sim
\frac{\sigma}{h\kappa^2} \sim \frac{\sigma h^2}{V^2h} \sim
\frac{M_{\mathrm{d}}}{M_{\mathrm{t}}}.
\end{equation}

Here $M_{\mathrm{d}} \sim \sigma h^2$ and $M_{\mathrm{t}} \sim
V^2h$ are the mass of a disk and the total mass of a galaxy,
respectively, within the fixed radius (in the units of $h$). The
thinnest galaxies can therefore be expected to be those with the
highest mass fraction of the spherical halo. This conclusion
agrees well with the results of the 3D $N$-body numerical
simulations of collisionless disks (Zasov and Morozov 1985; Zasov
et al. 1991; Mikhailova et al. 2001).

When  applied to real galaxies the situation may be complicated
by a different factors which can lead to the increasing the
thickness of quasi-equilibrium disks in the process of their long
evolution due to slow growth of velocity dispersion (Gerssen et
al. 2000; Binney 2000). These factors include the scattering of
disk stars during their interaction  with giant molecular clouds
or globular clusters; interaction of stars with density waves;
merging of small satellites, which could cross repeatedly over the
disk, star formation in the process of gas accretion onto the
disk, which has not yet reached equilibrium, and gravitational
perturbations due to neighboring galaxies. The latter effect
shows up conspicuously in the fact that the relative thickness of
disks in interacting systems is about twice larger than in
galaxies without close neighbors (Reshetnikov and Combes 1997).

Note that the efficiency of all the processes mentioned above
should be different at different galactocentric distances, whereas
photometric measurements of edge-on galaxies imply that disk
thickness varies only slightly with radius (van der Kruit and
Searle 1981a, b; Barnaby and Thronson 1992). (Note however, that
some galaxies appear not to obey this rule -- see de Grijs and
Peletier 1997).  The conclusion about the disk thickness
remaining constant over a wide interval of galactocentric
distances also follows from numerical $N$-body simulations of the
dynamical evolution of initially cold (along the $z$-coordinate)
collisionless disks (Mikhailova et al. 2001).

To clarify the processes that determine the vertical scale height
of a stellar disk, it is worth verifying whether the relative
thickness of disks seen edge-on correlates with the dark halo
mass, and this is just the aim of this work.

$$
$$

\textbf{GALAXY SAMPLES USED}

We chose the galaxies satisfying the condition $a/b\geq7$ in $B$
band, which is the underlying criterion of the Flat Galaxy
Catalog (RFGC, Karachentsev et al. 1999). The objects obeying this
criterion are mostly Sc--Sd galaxies ($\sim75\%$). These are
disk-dominated galaxies with the small bulge contribution to the
integrated luminosity (although in some cases the bulge presence
is clearly seen in central regions), making it easier to
determine their vertical and radial scale lengths and the total
disk luminosities.

In this work we use two samples of edge-on galaxies. The first
sample (below we will refer to it as BTA sample) includes  121
late-type galaxies of Karachentsev et al.' Catalog. For these
galaxies $R$-band surface CCD photometry was performed at BTA
telescope (Karachentsev et al. 1992). We excluded from the initial
sample the objects with uncertain shapes of their outer isophotes
and those with isophotal asymmetry in the inner region, which
might indicate that the disk inclination differs appreciably from
$90^{\circ}$. Nearby galaxies ($V < 750~km/s$), Virgo members,
and galaxies with large galactic extinction ($A_R
> 0.5$) were also excluded. Our final analysis was based on the final
sample of 51 galaxies.

Karachentsev et al. (1992) gave the estimates of $R$-band axial
ratios $a/b$ , angular sizes of the semi-major axes of the 23 and
$24^m/\mathrm{arcsec}^2$ isophotes parallel with the corresponding
isophotal magnitudes, and photometric profiles of the observed
galaxies.

We estimated the radial scale length $h$ by fitting the
photometric major-axis profile to a function implied by relation
(2) at $z_0$ = 0. Given $h$, the vertical scale height $z_0$ can
be determined by measuring the semi-major ($a$) and semi-minor
($b$) axes of a certain isophote of the galaxy (sufficiently far
from the center to minimize bulge effects) and using relation
(2). The latter implies for the points lying along the major
($a,0$) and minor ($0,b$) axes:

\begin{equation}
\mathrm{sech}^2\left(\frac{b}{z_0}\right) = \frac{a}{h}
K_1\left(\frac{a}{h}\right).
\end{equation}

Unfortunately, the available photometric data were insufficient
to allow a more refined approach making use of the entire pattern
of the two-dimensional brightness distribution of a galaxy. We
found the galaxies of our sample to have median isophotal major
axis to the radial disk scale length ratios, $a/h$, of 2.9 and
3.7 for the isophotes $23^m / arcsec^2$ and $24^m / arcsec^2$,
respectively.

As a second sample we analyzed 60 RFGC galaxies chosen for their
largest angular size, whose vertical ($z_0$) and radial ($h$)
disk scale lengths could be determined in a more rigorous way --
by modeling photometric cross sections along and across the major
axis of the galaxy. For this purpose we used the 2MASS survey
$K_{\mathrm{s}}$-band near-infrared images available from NASA
Extragalactic Database (NED). A detailed description of the
procedure we used to determine the photometric parameters is
given by Bizyaev and Mitronova (2002). The above authors obtained
the vertical disk scale heights averaged over 20--30 vertical
cross sections. We fitted each profile to a $I =
I_0\mathrm{sech}^2(z/z_0)$ law with allowance for atmospheric
blurring.

The radial disk scale lengths were determined from the cross
sections parallel to the major axis of the galaxy (but not closer
than $\sim2''$ to avoid the dustiest regions). To minimize bulge
effects in the estimated photometric disk parameters $z_0$ and
$h$, we also excluded the centermost regions in the cases where
the isophotal ellipticity decreased centerward (due to the bulge).
We treated $z_0$ and $h$ in equation (2)
and the central brightness of the exponential disk as free
parameters of the photometric model.

\begin{figure}
\epsfxsize=9cm {\epsfbox{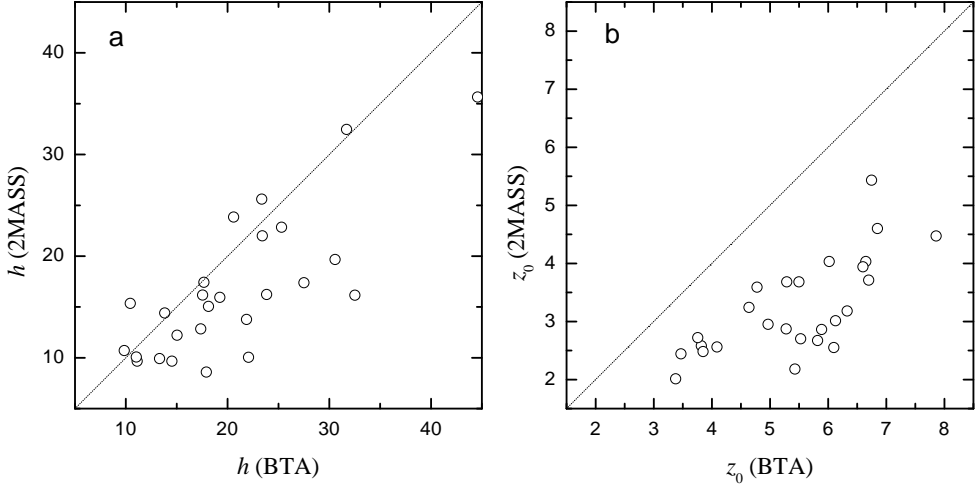}} \caption{A comparison of
radial (a) and vertical (b) disk scale lengths estimations
obtained by the different methods in different color bands (BTA
-- $R$ band; 2MASS -- $K_{\mathrm{s}}$ band) for the galaxies
common for both samples. \hfill}
\end{figure}

\begin{figure}
\epsfxsize=9cm \epsfbox{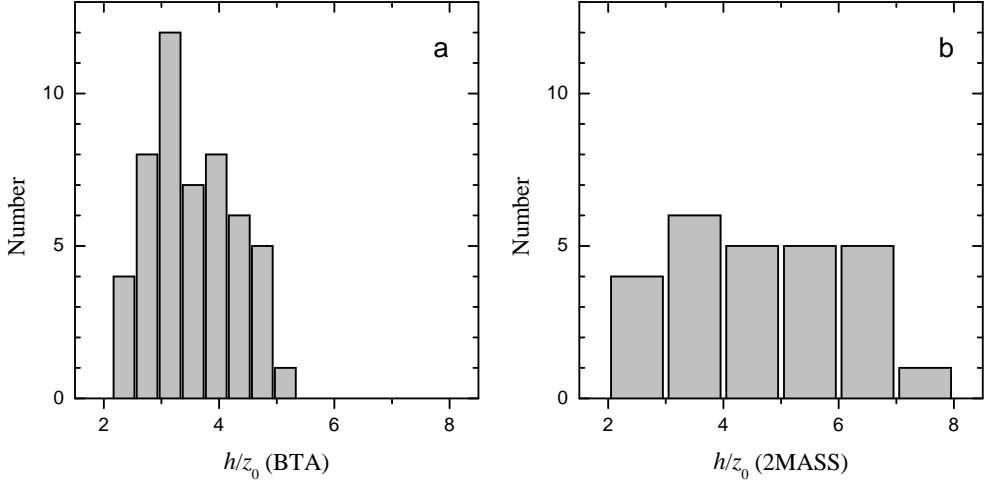} \caption{ Histogram of the
radial-to-vertical disk scale length ratio, $h/z_0$, for the
galaxy samples considered: (a) -- $R$-band (BTA sample); (b)
$K_s$-band (2MASS sample). \hfill}
\end{figure}

For a comparative analysis of disk scales of two samples we
selected 24 brightest ($K_s<10.5^m$) and relatively distant ($V$
> 750~km/s) galaxies of the second sample excluding the objects
with the strongest galactic extinction $A_K
> 0.25$ and probable
Virgo cluster members.

The first (BTA) sample of galaxies with $R$-band photometry and
the initial sample of galaxies from 2MASS  catalog have 28
objects in common. When comparing the two samples we excluded two
objects with supposedly non-exponential profiles yielding strongly
discrepant scale length estimates obtained in two samples (UGC
542 = RFGC 206 and UGC 7774 = RFGC 2336). In Fig.~2 we compare
the independently determined $z_0$ and $h$ (in acrseconds). The
median radial and vertical scale height ratios for both samples
are: $h(BTA)/h(2MASS)=1.21 \pm 0.08$ and
$z_0(BTA)/z_0(2MASS)=1.66 \pm 0.07$, respectively. The relation
between the radial scale lengths agrees well with the conclusion
of de Grijs (1998) that near-infrared ($K$) photometric scale
lengths are systematically smaller than those measured at shorter
wavelengths (by a factor of about $\approx 1.2$ and $\approx 1.6$
compared to the $I$- and $B$-band data, respectively). The scale
length ratio $z_0/h$ also decreases as one passes to longer
wavelengths (see Fig.~6 of de Grijs (1998)). According to our
measurements, the relative thickness of the galaxies of the first
sample ($R$-band) is also greater than that of the second  sample
($K_s$-band) (Fig.~3): the mean ($h/z_0$) are equal to $3.52 \pm
0.1$ and $4.93 \pm 0.34$ for BTA and 2MASS samples, respectively.
This effect, however, can be partially due to systematically
overestimated $z_0$ based on BTA data, because the  method
employed is sensitive to the eventual bulge effects in the
computed minor axes of the isophotes used to determine the
vertical scale height.

Our photometric measurements showed that the galaxies of the BTA
sample have a mean integrated color index of $B-R=1.06 \pm 0.05$.
The mean color indices of the edge-on galaxies common for the two
samples are $B-K_s = 3.34 \pm 0.17$ and $R-K_s = 2.24 \pm 0.12$,
respectively. These results agree well with the integrated colors
of 86 Sc--Sd galaxies oriented almost face-on (de Jong 1996). The
absence of strong reddening of edge-on galaxies is no surprise: a
dust lane extending along their major axis can strongly decrease
the observed luminosity, while having little effect on the color
if the optical depth of the dust $\tau >> 1$.

$$
$$

\textbf{VERIFICATION OF THE DEPENDENCE OF THE RELATIVE DISK
THICKNESS ON THE $M/L$ RATIO OF A GALAXY}

If the above mentioned assumption about the decrease of the
relative disk thickness with the mass fraction of the spherical
component (dark halo) is true, \footnote{The samples considered
consist mostly of late-type galaxies without massive bulges and
therefore the bulk of the mass of the spherical component belongs
to the dark halo indeed } one should expect the $z_0/h$ ratio to
be the lowest in galaxies with high ratio of the integrated mass
to the integrated red (infrared) luminosity: the latter is
sensitive only slightly to the ongoing star formation and
therefore better than the blue light corresponds to the total
mass of the stellar population of the disk.

Hereafter we determine the masses of galaxies inside fixed radius
of $R_m = 4h$ within which the luminosity (actually, the disk
luminosity) was determined from photometric data. Beyond $4h$
less than $10\%$ of the mass of the exponential disk is located
-- even in the absence of the usually observed steepening of the
radial distribution at large $R$. We assume that the total mass
$M_{\mathrm{t}}$ of the galaxy within $R_m$ is approximately
equal to $W_H^2R_m/4G$, where $W_H$ is the HI line width.
\footnote{In this work we use the $W_{50}$ width at $50\%$ of the
maximum (adopted from LEDA database). However, the choice between
$W_{20}$ and $W_{50}$ is of no fundamental importance, because
both quantities are close to twice the maximum velocity of gas
rotation.} This simple expression for mass is, strictly speaking,
correct for spherically-symmetrical systems, however, this
assumption introduces a rather small error. Numerical simulations
of galaxies with the measured velocity dispersion of the old disk
stellar population imply halo masses exceeding significantly the
disk masses within chosen $R_m$ in most of the cases (see for
example Zasov et al. 2000; Khoperskov et al. 2001). However, even
if the mass of the halo  is equal to that of the thick disk
within $R_m = 4h$, the above formula overestimates the mass
$M_{\mathrm{t}}$ only by $\approx 25\%$.

For the galaxies of the first sample we estimated the galaxy
luminosities from the $24^m/\mathrm{arcsec}^2$ isophotal
magnitudes extrapolating them if necessary out to $R_m$ based on
the radial brightness scale length. The luminosity of galaxies of
the second sample were restored from their  photometric model
parameters. If the relative masses of disk and halo are unknown,
the mass of the disk cannot be estimated from the rotation
velocity and therefore we infer it from the disk luminosity
assuming that $M_{\mathrm{d}} =
A(\lambda)L_{\mathrm{d}}\times(M/L)_{\mathrm{d}}$, where
$A(\lambda) > 1$ is the factor that allows for internal
extinction (which is important for the $R$ band);
$L_{\mathrm{d}}$, the observed disk luminosity, and
$(M/L)_{\mathrm{d}}$, the integrated mass-to-luminosity ratio of
the stellar population in the chosen photometric band. The
total-to-disk mass ratio can therefore be written in the
following form:
\begin{equation}
\frac{M_{\mathrm{t}}}{M_{\mathrm{d}}} \approx  \frac{W_{50}^2
h}{A(\lambda)GL_{\mathrm{d}}(M/L)_{\mathrm{d}}}.
\end{equation}

\begin{figure}
 \epsfxsize=9cm {\epsfbox{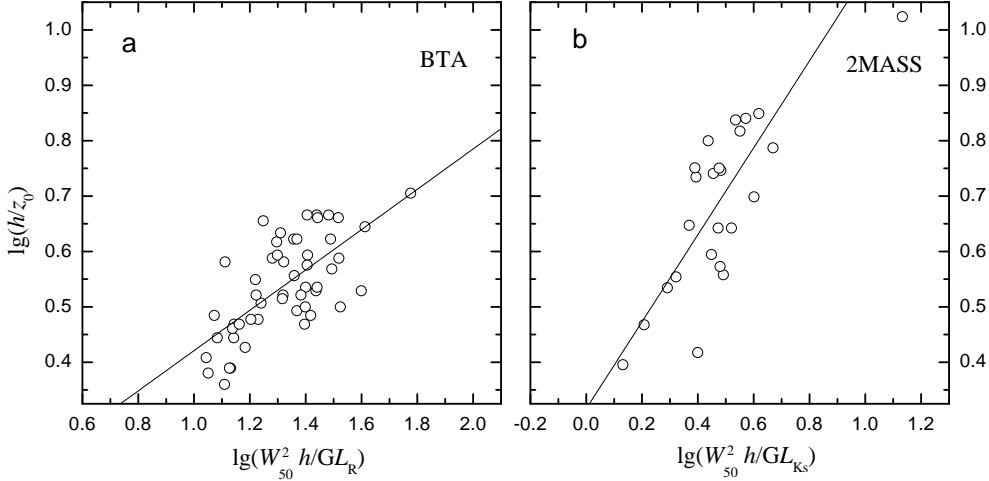}}
\caption{Relation between the photometrically determined radial
-to - vertical disk scale length ratio  and the quantity
$W^2_{50}h/GL_{(R,Ks)}$, which determines the $M_{\mathrm{t}} /
L_{\mathrm{d}}$ ratio within $R = 4h$: (a) $R$-band (BTA data);
(b) $K_{\mathrm{s}}$-band (2MASS data). \hfill}
\end{figure}

The luminosity underestimation  of an edge-on galaxy is difficult
to take into account: it can be important even in the infrared.
The extinction correction applied to reduce the $R$-band
magnitudes of edge-on galaxies to those of face-on galaxies
exceeds, on the average, $1^m$ (Tully et al. 1998). The
estimation of $M_d$ from the observed luminosity is further
complicated by the large scatter of coefficient $A(\lambda)$,
which, in turn, can depend on the disk mass and thickness. For
the second galaxy sample the photometric estimates should be much
less affected by dust. The reasons for this are twofold: (1)
$K_s$-band extinction in galaxies resulting from their edge-on
orientation does not, on the average, exceed $0.3^m$ (Tully et
al. 1998), and (2) when estimating the scale lengths we excluded
the regions close to the Galactic plane, which suffer from the
strongest extinction. However, in spite of the simplifying
assumptions adopted here both the first and the second galaxy
samples exhibit conspicuous relations between $h/z_0$ and
$M_t/L_d$ (or, to be more precise, a quantity proportional to
this ratio)
--- see Fig.~4 -- with the correlation coefficients equal to 0.68
and 0.73, for the first and the second samples respectively. This
relation, which corroborates the conclusion that the disk
thickness decreases with the relative mass of the spherical
component is the main result of this work.

The scatter of data points on the diagrams shown here is due to
the errors in the estimates of the parameters used, the
difference of $A_\lambda$ and $M/L_d$ of the stellar populations
of individual galaxies, and unaccounted physical factors, which
may increase the disk thickness (see Introduction). The
differences between the slopes based on two galaxy samples must
be real despite the uncertainty of the inferred slope of the
relation in Fig.~4(a) (photometric estimates based on 2MASS data
are more reliable): the shallower behavior of the $R$-band
relation agrees qualitatively with the fact that thinner galaxies
(in the upper part of the diagram) are more extinction affected
and thus have their $M/L_d$ overestimated.

\begin{figure}
\epsfxsize=9cm \epsfbox{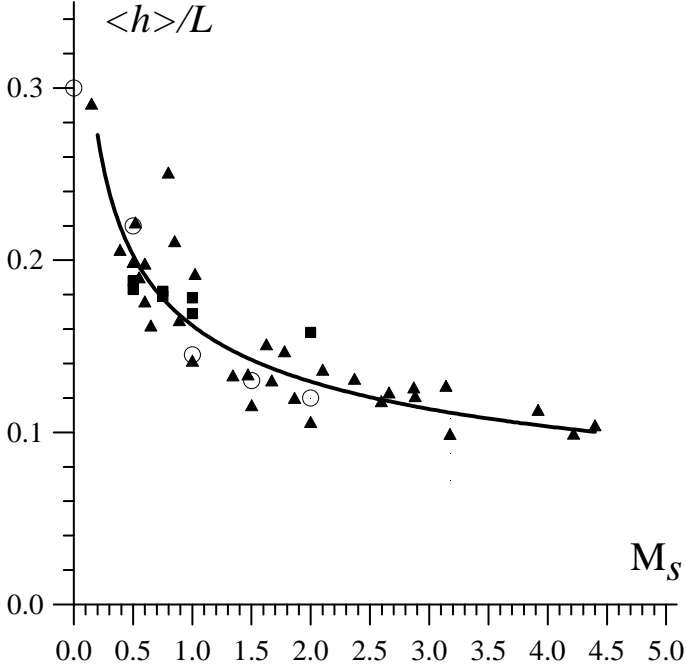} \caption{Relation between
the rms distance $\langle h\rangle$ of points from the disk plane
expressed in the units of the radial disk scale length and the
halo to disk mass ratio, $M_s$, obtained from $N$-body numerical
simulations of galaxies with marginally stable disks. The figure
is adopted from  Mikhailova et al. (2001, Fig.~2). In our
notations $\langle h\rangle/L \approx z_0/h$. \hfill}
\end{figure}

$$
$$

\textbf{DISCUSSION AND CONCLUSIONS}

This study may be the first to show that the expected relation
exists between the stellar disk thickness and the relative masses
of spherical and disk components of a galaxy. Note that the
conclusion that marginally stable collisionless disks become
thinner with increasing mass fraction of their spherical
components (in the absence of external gravitational
perturbations) was first reached from $N$-body numerical
simulations of three-dimensional disks in a fixed  field of the
spherical component starting from an unstable state with low
vertical velocity dispersion (Zasov et al. 1991; Mikhailova et
al. 2001). As simulations showed, the velocity dispersion $C_z$
increases from the initial values reaching a certain level
(decreasing with $R$) during the time interval equal to several
rotation periods at the outer disk edge - evidently as a result
of the development of bending perturbations. Eventually the disk
becomes  marginally stable against both perturbations in its
plane and the bending perturbations. Here we refer to Khoperskov
et al. (2001, 2002) for a detailed description of numerical
simulations.

Figure~ 5,  taken from the paper by Mikhailova et al, 2001
(their  Fig.~2) compares the relative disk thickness and
$M_s=(M_t-M_d)/M_d$ -- the spherical-to-disk mass ratio -- based
on the results obtained by constructing numerical models for
galaxies with different component masses and different shapes of
rotation curves corresponding to those actually observed in real
galaxies.

To compare the observed and model relations shown in Figs.~4 and
5, one must convert disk luminosities into disk masses. Assuming,
like we did in the previous section, that $W_{50}^2h/G$
determines the total mass of the galaxy within $R_m = 4h$, we can
write the quantity $M_s$ laid off along the horizontal axis in
Fig.~5 as:
\begin{equation}
 M_s = \frac{W_{50}^2h}{G\times M_d}-1,
\end{equation}
It follows from this equation that:
\begin{equation}
\frac{W_{50}^2h}{G\times L_d}  = (M_{\mathrm{s}}+1)(M/L)_d,
\label{e:W_50}
\end{equation}
where $(M/L)_{\mathrm{d}}$ is the disk mass-to-luminosity ratio
for the chosen spectral interval.

\begin{figure}
\epsfxsize=9cm \epsfbox{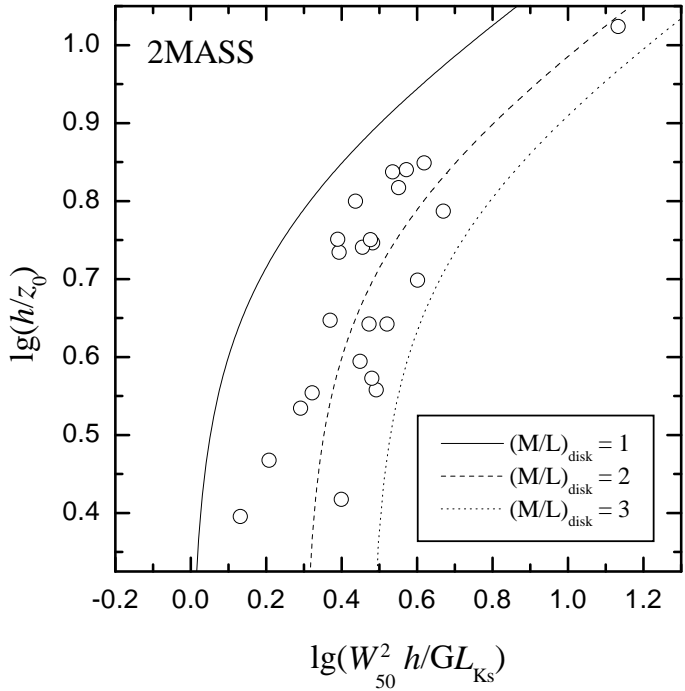} \caption {The same diagram as in
Fig.~4b with superimposed collisionless-disk relation inferred
from numerical simulations (the curve in Fig.~5) recalculated for
three disk mass-to-luminosity ratios (in solar units):
$(M/L)_d=1$, 2, and 3. \hfill}
\end{figure}

Figure~6 shows in a logarithmic scale the diagram given in Fig.~4
with the superimposed curve from Fig.~5 computed in accordance
with equation (7) for three mass-to-$K_s$-band luminosity ratios
($M/L_{\mathrm{d}} = 1$, 2, and 3).

Evolutionary models yield for the stellar population of
cosmological-age galaxies a mass-to-luminosity ratio of
$(M/L)_{\mathrm{model}}\approx 1$ for the photometric $K$ band,
which is close to $K_{\mathrm{s}}$ (Bell and de Jong 2001). This
ratio remains somewhat uncertain due to the lack of data about the
low-mass end of the stellar mass function. All galaxies are
actually situated in the domain between the adopted ratios, which
are quite reasonable for an old stellar population. It shows that
models of marginally stable disks agree well with observations.
This leads us to conclude that for most of the galaxies the
mechanisms of additional disk heating (scattering by  massive
clouds, tidal perturbation of the disk) are not crucial for the
formation of the vertical disk structure. Hence the approximately
constant disk thickness along the radius may be considered as a
result of the two opposite tendencies influencing the disk
thickness:
--- the radial decrease of surface disk density and the decrease
of the velocity dispersion at which the disk reaches stable
equilibrium -- both factors cancel almost exactly each other.

\begin{figure}
\epsfxsize=9cm {\epsfbox{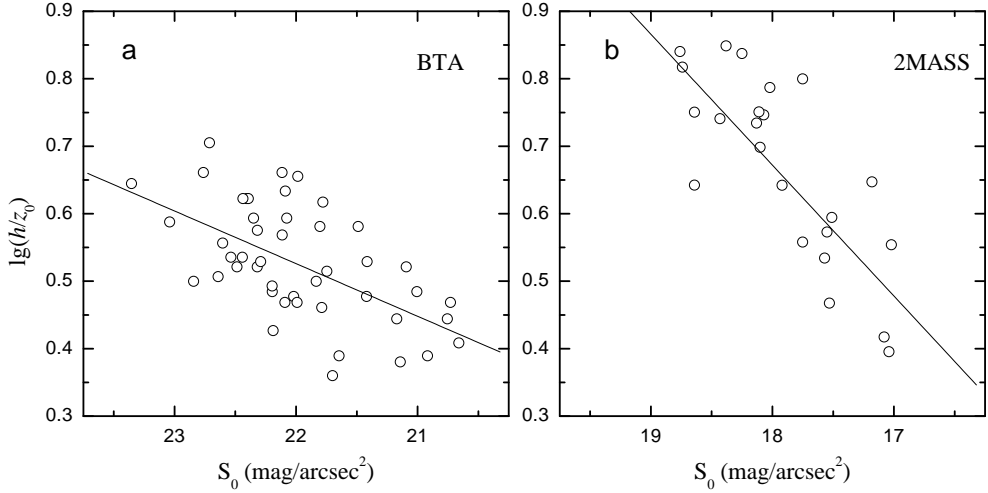}} \caption{Radial-to-vertical
disk scale length ratio as a function of deprojected central disk
brightness in magnitudes in the (a) $R$- and (b)
$K_{\mathrm{s}}$-band. \hfill}
\end{figure}

Figure~7 compares the relative disk thickness with central
surface brightness (in magnitudes) reduced to face-on position
using model $R$- and $K_{\mathrm{s}}$-band brightness
distributions. The correlation between these parameters is even
more conspicuous than that between $z_0/h$ and
$(M/L)_{\mathrm{t}}$, although the very existence of such a
relation is nothing unexpected:  "normal" and low surface
brightness galaxies were already shown to exhibit a close
relation between $S_0$  and integrated ratio $(M/L)_{\mathrm{t}}$,
which characterizes the dark halo mass fraction (MacGaugh and de
Block 1998). The lower the central surface brightness (and,
consequently, the surface density), the higher the dark halo mass
fraction within the chosen $R = R_m$. The correlation between
these two quantities implies, in particular, the existence of a
single linear (in the logarithmic terms) Tully--Fisher relation
(luminosity or mass of the disk -- rotation velocity) for
galaxies with different surface brightness $S_0$ (MacGaugh and de
Block 1998).

This relation manifests itself most conspicuously in the diagram
shown by Bizyaev and Mitronova (2002), which is based on an
analysis of a 153 galaxy sample from the 2MASS survey. This
relation appears to be more scattered at  longer-wavelength
bands  (as is evident from a comparison of diagrams a and b in
Fig.~7); Bizyaev and Kaisin (in preparation) and Bizyaev (2000)
came to the same  conclusion based on $R$ and $I$-band photometry,
respectively. The same dependence in the $B$-band is the least
conspicuous (see Fig.~9 in the paper of de Grijs (1998)). The
differences between the correlation coefficients and slopes of
relations shown in Fig.~7 are evidently due to selective internal
extinction, which is strongest in galaxies with thin disks and
becomes more important at shorter wavelengths. Indeed, the
underestimated brightness (or overestimated $S_0$, expressed in
magnitudes) for galaxies with "thin" disks result in an
underestimated slope of the relation in Fig.~7. It is not
surprising that the relation is more conspicuous in the
$K_{\mathrm{s}}$ band (Fif.~7b) where the  internal extinction
amounts only to several tenths of a magnitude.

Thus the relations obtained lead us to conclude that the thinnest
edge-on galaxies are to be (after deprojection to a face-on
orientation) low surface brightness spirals whose observed
brightness is enhanced by projection effect and whose dark-halo
mass exceeds significantly the mass of the disk.

Note, however, that not all disk galaxies appear to obey the
"dark-halo mass -- disk thickness" relation. Relatively thick
disks are observed not only in interacting systems (Reshetnikov
and Combes, 1997) but also in comparatively low luminous Irr
galaxies, for which some authors pointed out a deficit of systems
with strong apparent flattening (Hodge and Hitchcock 1966; Thuan
and Seitzer 1979; van den Bergh 1988), and all this in spite of
the fact that the dark halo mass fraction in low luminosity
galaxies is, on the average, higher  than in galaxies with high
luminosity (Persik and Salucci 1996 ; Ashman 1992; Cote et al.
2000). Relatively thick disks observed in Irr-galaxies may be a
result of, among other things, a certain threshold level of
stellar velocity dispersion, which cannot be lower than the
velocity dispersion of gaseous clouds (usually $\approx10$~km/s).
Unfortunately, the  strong contribution of young stars to the
disk luminosity in Irr-galaxies and their nonuniform distribution
within the galaxy complicate the vertical photometric disk
structure and increase the uncertainty of the photometric tilt
estimates compared to what we have in the case of spiral
galaxies, thus preventing any direct comparison of the stellar
disk thickness of these two types of objects.

\newpage
\textbf{ACKNOWLEDGMENTS}

This work was supported by the Russian Foundation for Basic
Research (grant no.~01-02-17597). The authors thank
I.D.Karachentsev for sharing the original data on galactic
photometry.

\bigskip

{REFERENCES}

K.M.Ashman, Publ.Astron.Soc.Pac., 104, 1109 (1992).

D.Barnaby and H.A.Thronson, Astron.J., 103, 41 (1992).

E.F.Bell and R.S.de Jong, Astrophys.J., 550, 212 (2001).

J.J.Binney, Astron.Soc.Pac.Conf.Ser., 197, 107 (2000).

D.V.Bizyaev, Pis'ma Astron.Zh., 26, 266 (2000) [Astron.Lett. 26, 219 (2000)].

D.V.Bizyaev and S.S.Kaisin (in preparation).

D.V. Bizyaev and S.N. Mitronova, Astron. and Astrophys., 389, 795 (2002)

R.Bottema, Astron. and Astrophys., 275, 16 (1993)

S.Cote, C.Carignan, and K.C.Freeman, Astron.J, 120, 3027 (2000)

R.de Grijs, Mon. Not. Roy. Astron. Soc., 299, 595 (1998)

R.de Grijs and R.F.Peletier, Astron. and Astrophys., 320, L21 (1997)

R.de Grijs and P.C.van der Kruit, Astron Astrophys, Suppl.
Ser., 117, 19 (1996)

R.de Grijs, M.Kregel and K.H.Wesson, Mon.Not.Roy.Astron.Soc., 324, 1074
(2001)

R.S.de Jong, Astron. and Astrophys., 313, 377 (1996)

J.Gerssen, K.Kuijken, and M.R.Merrifield,
Mon.Not.Roy.Astron.Soc., 317, 545 (2000)

P.W.Hodge and J.L.Hitchcock, Publ.Astron.Soc.Pac., 78, 79 (1966)

I.D.Karachentsev, Ts.B. Georgiev, S.S.Kajsin, et al.,
Astron. and Astrophys.Trans.2, 265 (1992).

I.D.Karachentsev, V.E.Karachentseva, Yu.N. Kudrya, and
S.L.Parnovski, Pis'ma Astron.Zh . 23, 652 (1997)
[Astron.Lett. 23, 573 (1997)].

I.D.Karachentsev, V.E.Karachentseva, Y.N. Kudrya et al.,
Bull.SAO 47, 5 (1999).

A.V.Khoperskov, A.V. Zasov, N.V.Tyurina, Astron.Zh. 78, 213 (2001)
[Astron.Rep., 45, 180 (2001)].

Yu.N.Kudrya, I.D.Karachentsev, V.E.Karachentseva and
S.L.Parnovski,  Pis'ma Astron.Zh., 20, 13 (1994) [Astron.Lett., 20, 8 (1994)].

J.Ma, J.L.Zhao, C.G.Shu and Q.H.Peng, Astron. and Astrophys., 350,
31 (1999).

J.Ma, Q.-H.Peng, R.Chen, Z.-H.Ji Astron. and Astrophys. S., 126,
503 (1997).

S.McGaugh and W.de Block, Astrophys.J., 499, 41 (1998).

S.McGaugh, J.M.Schombert, G.D.Bothun and W.J.G.de Blok,
Astrophys.J., 533, 95 (2000).

E.A.Mikhailova, A.V.Khoperskov, and S.S.Sharpak, in "Stellar
Dynamics - from Clasic to Modern", Ed.by L.P.Ossipkov and

I.I.Nikiforov (St.Petersburg State University, St.Petersburg,
2001), p.147.

M.Persic and P.Salucci, Mon.Not.Roy.Astron.Soc., 281, 27 (1996).
V.L.Polyachnko and I.G.Shukhman, Pis'ma Astron.Zh., .3, 254
(1977) [Sov.Astron.Lett.3, 134 (1977)].

V.Reshetnikov and F.Combes, Astron. and Astrophys. 324, 80
(1997).

T.X.Thuan and P.O.Seitzer, Astrophs.J., 231, 680 (1979).

R.B.Tully, M.J.Pierce, Jia Sheng Huang, et al.,
Astron.J., 115, 2264 (1998).

P.C.van der Kruit and L.Searle, Astron. and Astrophs., 95, 105
(1981a).

P.C.van der Kruit and L.Searle, Astron. and Astrophs., 95, 116 (1981b).

A.V.Zasov and A.G.Morozov, Astron.Zh., 62, 475 (1985)
[Sov.Astron.Lett.29, 277 (1985)].

A.V.Zasov, D.I.Makarov, and E.A.Mikhailova, Pis'ma
Astron.Zh., 17, 884 (1991) [Sov.Astron.Lett. 17, 374 (1991)].

A.V.Zasov, A.V.Khoperskov and N.V.Tyurina,
in "Stellar Dynamics - from Classic to Modern", Ed. by L.P.Ossipkov
and I.I.Nikiforov (St.Petersburg State University, St.Petersburg,
2001), p.95.

\end{document}